\documentclass[apj]{emulateapj}
\slugcomment{Accepted for publication in ApJ 2010 May 23}

\begin{document}


%%

%\title{ Multiband Optical Short Term Variability of Low$-$Energy Peaked Blazars}
\title{Detection of Intra-day Variability Timescales of Four High Energy Peaked Blazars
with XMM$-$Newton}

\author{Haritma Gaur\altaffilmark{1}, Alok C.\ Gupta\altaffilmark{1}, Pawel
Lachowicz\altaffilmark{2, 3}, Paul J. Wiita\altaffilmark{4,5}}

\altaffiltext{1}{Aryabhatta Research Institute of Observational Sciences (ARIES),
Manora Peak, Nainital - 263129, India}
\altaffiltext{2}{Nicolaus Copernicus Astronomical Center, Polish Academy of Sciences,
ul.\ Bartycka 18, 00-716 Warszawa, Poland}
\altaffiltext{3}{Center for Wavelets, Approximations and Information Processing, Temasek
Laboratories, National University of Singapore, 5A Engineering Dr 1, \# 09-02 Singapore 117411}
\altaffiltext{4}{Department of Physics and Astronomy, Georgia State University, P.O.\ Box 4106,
Atlanta, GA 30302-4106}
\altaffiltext{5}{Department of Physics, The College of New Jersey, P.O.\ Box 7718, Ewing, NJ 08628}

\email{haritma@aries.res.in, acgupta30@gmail.com, pawel@ieee.org, wiita@chara.gsu.edu}

%{Phone No. +91$-$9457623544, Fax No. +91$-$5942$-$233439}
%

%\altaffiltext{1}{National Astronomical Observatories/Yunnan Observatory, Chinese
%Academy of Sciences, P.O. Box 110, Kunming, Yunnan 650011, China.}
%\altaffiltext{1}{Visiting Astronomer, Cerro Tololo Inter$-$American Observatory.
%CTIO is operated by AURA, Inc.\ under contract to the National Science
%\altaffiltext{2}{Society of Fellows, Harvard University.}
%\altaffiltext{3}{present address: Center for Astrophysics,
%    60 Garden Street, Cambridge, MA 02138}
%\altaffiltext{4}{Visiting Programmer, Space Telescope Science Institute}
%\altaffiltext{5}{Patron, Alonso's Bar and Grill}

\begin{abstract}

We selected a sample of 24 XMM-Newton light curves  (LCs) of four high energy peaked blazars,
PKS 0548$-$322, ON 231, 1ES 1426$+$428 and PKS 2155$-$304. These data comprise continuous light  
curves of 7.67h to 18.97h in length. We searched for possible quasi-periodic oscillations 
(QPO) and intra-day variability (IDV) timescales in the LCs of these blazars. We found a likely 
QPO in one LC of PKS 2155$-$304 which was reported elsewhere (Lachowicz et al.\ 2009).  
In the remaining 23 LCs we found  hints of possible weak QPOs in one LC of each of ON 231 and 
PKS 2155$-$304, but neither is statistically significant. We found IDV timescales that 
ranged from 15.7 ks to 46.8 ks in 8 LCs. In 13 LCs any variability timescales were longer than 
the length of the data. Assuming the possible weak QPO periods in the blazars PKS 2155$-$304 
and ON 231 are real and are associated with the innermost portions of their accretion disk, 
we can estimate that their central black hole masses exceed 1.2 $\times$ 10$^{7}$ M$_{\odot}$. 
Emission models for radio-loud active galactic nuclei (AGN) that could explain our results are 
briefly discussed.  

\end{abstract}


%% subject header.  Objects should be in the appropriate "individual"
%% headers (e.g. quasars: individual, stars: individual, etc.) with the
%% additional provision that the total number of headers, including each
%% individual object, not exceed six.  The \objectname{} macro, and its
%% alias \object{}, is used to mark each object.  The macro takes the object
%% name as its primary argument.  This name will appear in the paper
%% and serve as the link's anchor in the electronic edition if the name
%% is recognized by the data centers.  The macro also takes an optional
%% argument in parentheses in cases where the data center identification
%% differs from what is to be printed in the paper.

\keywords{galaxies: active $-$ BL Lacertae objects: general $-$ BL Lacertae objects: 
individual (\object{PKS 0548$-$322}; \object{ON 231}; \object{1ES 1426$+$428}; 
\object{PKS 2155$-$304})}

%\keywords{AGN: blazar: optical: observations $-$ blazars: individual: S5 0716$+$714}
%(\objectname{NGC 6397},
%\object{NGC 6624}, \objectname[M 15]{NGC 7078},
%\object[Cl 1938$-$341]{Terzan 8})}


\section{Introduction}

Blazars, including BL Lacertae objects (BL Lacs) and flat spectrum radio quasars (FSRQs), are
extragalactic radio sources with relativistic jets aligned nearly ($\lesssim$ 10$^{\circ}$) with the
line of sight (e.g., Urry \& Padovani 1995). Blazar emission extends across the 
entire electromagnetic (EM) spectrum, is predominantly nonthermal, and shows significant polarization 
in the radio and visible bands where it can be measured.  Blazars show detectable 
flux variations on diverse timescales ranging from a few  minutes through days and months to decades  
through all EM bands. Blazar variability timescales have often
been somewhat arbitrarily divided into three classes: timescales from minutes to less than a 
day are called intra-day variability (IDV); those from several days to a few months are known as 
short timescale variability (STV); while long term variability (LTV) covers changes from several 
months to many years (e.g., Gupta et al.\ 2004).

Short and intra-day X-ray variability has been investigated in
various classes of AGNs (e.g., Edelson \& Nandra 1999; Uttley et al.\ 2002; Markowitz et al.\
2003; Vaughan
et al.\ 2003b; McHardy et al.\ 2004, 2005; Espaillat et al.\ 2008; Lachowicz et al.\
2009; and references therein). The great majority of these studies have been done for the brighter 
sources, which are typically nearby Seyfert galaxies, and were  reviewed recently by Uttley (2007). 
There were claims of the detection of QPOs in a few AGNs on the short and IDV
timescales in early X-ray 
observations (Fiore et al.\ 1989; Papadakis \& Lawrence 1993; Iwasawa et al.\ 1998).
But all of these claimed QPO
detections  were later found to be not statistically significant
(Tagliaferri et al.\  1996; Benlloch et al.\ 2001; Vaughan 2005; Vaughan \& Uttley 2006). 
There have, however, been few stronger  claims 
of QPO detection recently in X-ray data of various classes of AGNs. The first significant
detection of a short X-ray 
QPO of $\sim$ 1 hour timescale has been reported for RE J1034$+$396, which is a narrow
line Seyfert 1  galaxy (Gierli{\'n}ski et al.\ 2008). Espaillat et al.\ (2008) have reported an X-ray
QPO on the timescale 
of 3.3 ks in 3C 273, a FSRQ. Very recently, Lachowicz et
al.\ (2009) have
reported a detection of a X-ray QPO in the BL Lac PKS 2155$-$304 on a timescale of $\sim$
4.6 hours. All 
three of these QPO detections on IDV timescales were based on observations made with
XMM-Newton. By using All Sky Monitor data from the Rossi X-ray Timing Explorer,
 Rani et al.\ (2009) have reported a probable QPO from
 the BL Lac  AO 0235$+$164 on a STV timescale of $\sim$ 18 days.  Such QPOs  may
shed new light on the physical processes at the source and associated X-ray
emission and so the search for their presence in the light curves (LCs) of AGN is important.
 
The full spectral energy distributions (SEDs) of blazars have double humped structures with each 
large spectral feature having comparable total powers.
 In these double humped SEDs, the first (lower energy) component of the SED is quite clearly dominated 
by synchrotron radiation from the relativistic jet and the second component is probably due to 
Inverse-Compton (IC) radiation. On the basis of their SEDs, blazars can be classified into two 
subclasses known as LBLs (Low-Energy Peaked Blazars) and HBLs (High Energy Peaked Blazars), 
though those with intermediate peaks are certainly found (Nieppola et al.\ 2006). In LBLs, the 
synchrotron component peaks at near-IR, optical or near-UV frequencies and the high energy component 
usually peaks
at GeV energies. In HBLs, the synchrotron component peaks somewhere in the X-ray band and the high energy 
component is seen to peak (or is extrapolated to do so) at TeV energies.

Here, we study four HBLs which are designated as TeV blazars because of their significant detections in 
that extreme energy band. Just  six years ago only six TeV blazars were known (see Krawczynski 
et al.\ 2004 for a summary of their properties).  Thanks to the development of 
new TeV facilities such as HESS (High Energy Stereoscopic System; Hofmann et al.\ 2003; Funk 
et al.\ 2004. Aharonian et al.\ 2006), MAGIC (Major Atmospheric Gamma-ray Imaging Cerenkov; 
(Baixeras et al.\ 2004; Cortina et al.\ 2005; Albert et al.\ 2006a) and VERITAS (Very Energetic 
Radiation Imaging Telescope Array System; Holder et al.\ 2006; Maier \ 2007, Maier et al.\ 2008)
there has been a complete revolution in TeV gamma-ray astronomy. These groups have
detected about a dozen new TeV emitting blazars (HBLs) (Albert et al.\ 2006b, 2007, 2008, 2009; 
Acciari et al.\ 2008, 2009a, 2009b;  Aharonian et al.\ 2007, 2008a, 2008b, 
2008c, 2008d, 2009a, 2009b, 2009c,  and references therein).  
The enlarged sample of TeV blazars will be very useful for our understanding of the emission mechanism of 
these extreme blazars through the study of their variability properties across the range of EM bands. Here, 
we report our search for X-ray variability on IDV timescales in four HBLs from the 
XMM-Newton satellite public archive. We plan to perform extensive 
variability studies of HBLs in other EM bands using ground based observations.

The motivation of this work is to examine the nature  of IDV in a
sample of four high energy peaked blazars, namely, PKS 0548$-$322, ON 231, 1ES 1426$+$428 and PKS
2155$-$304, in which synchrotron emission from the jets dominates up through the X-ray component. 
We will  use IDV data to search for QPOs (quasi-periodic oscillations) and 
variability timescales in the light curves (LCs) of these HBLs, recalling that the shortest 
timescales can give  
upper limits to the sizes of the emitting regions. When a blazar is  in a low flux state it is possible 
that an accretion disk related component of the X-ray emission is dominant and then any IDV timescale 
could give an estimate of the mass of the black hole presumed to reside at the center 
of the galaxy (e.g., Gupta et al.\ 2008, 2009). 
This possibility arises from the fact that some blazars show evidence of the ``big blue bump'' that is
almost certainly indicative of accretion disk emission (e.g., Raiteri et al.\ 2007) while in their 
low states, when the jets are weak or even absent.  Under these circumstances the X-ray corona 
expected above such disks and observed in radio-quiet AGN should be visible.  A break in the 
power spectral density (PSD) plot from 
intra-day LCs can also yield the black hole mass of the blazar (e.g., Uttley 2007). 
XMM-Newton IDV LCs also may possibly be used to make independent estimations of the  
Doppler factors of the blazars (Fan, Xie \& Bacon 1999). 

The paper is structured as follows.  In Section 2, we give brief descriptions of the data selection
criterion and the data reduction method. In Section 3, we discuss the techniques we used
to search for variability properties and provide the results in Section 4.  Our discussion is in Section 5 
and the conclusions  are given in Section 6. 

\section{XMM-Newton Data Selection and Processing}

\begin{table*}
{\bf Table 1. Observation log of XMM-Newton data for  high energy peaked blazars}
\scriptsize

\noindent
%\begin{center}
\begin{tabular}{lcccclccrr} \hline \hline
Source         & $\alpha_{2000.0}$ & $\delta_{2000.0}$ & $z$ & Date of Observation          & ~~~~~~PI       & GTI $^a$(ks) & Timescale(ks) & $\sigma_{frac}$ & $\mu$ (cts  s$^{-1})$  \\
               &                   &                   &     & dd.mm.yyyy                   &                &              &                 &  & \\\hline
PKS0548$-$322  & 05h 50m 41.0s & $-32^{\circ} 16^{'} 11^{\prime \prime}.0$ & 0.069  &  11.10.2002                 & Xavier Barcons & 45.5         &  Not Found       &2.65  &  13.82$\pm$0.03 \\
ON 231         & 12h 21m 31.7s & $+28^{\circ} 13^{'} 58^{\prime \prime}.5$ & 0.102  &  26.06.2002                 & Keith Mason    & 27.6         &  Weak QPO?$^b$   &31.58 &   1.89$\pm$0.06 \\
1ES1426$+$428  & 14h 28m 32.7s & $+42^{\circ} 40^{'} 21^{\prime \prime}.0$ & 0.129  &  16.06.2001                 & Bert Brinkman  & 62.0         & 46.8             &3.49  &  11.95$\pm$0.04 \\
               &               &                                           &        &  04.08.2004                 & Fred Jansen    & 61.9         & Not Found        &2.69  &  14.43$\pm$0.04 \\
               &               &                                           &        &  06.08.2004                 & Fred Jansen    & 68.3         & Not Found        &2.71  &  14.39$\pm$0.04 \\
               &               &                                           &        & 24.01.2005                 & Fred Jansen    & 29.8         & Not Found        &2.51  &  18.01$\pm$0.06 \\
               &               &                                           &        & 19.06.2005                 & Fred Jansen    & 37.9         & Not Found        &2.33  &  26.57$\pm$0.05 \\
               &               &                                           &        & 25.06.2005                 & Fred Jansen    & 33.9         & Not Found        &2.13  &  20.77$\pm$0.05 \\
               &               &                                           &        & 04.08.2005                 & Fred Jansen    & 39.9         & Not Found        &2.27  &  20.22$\pm$0.05 \\
PKS2155$-$304  & 21h 58m 52.1s & $-30^{\circ} 13^{'} 31^{\prime \prime}.1$ & 0.116  & 30.05.2000                 & Arvind Parmar  & 37.9         & 24.2             &2.89  &  54.76$\pm$0.05 \\
               &               &                                           &        & 31.05.2000                 & Arvind Parmar  & 59.2         & 41.2             &3.27  &  54.86$\pm$0.04 \\
               &               &                                           &        & 19.11.2000                 & Laura Maraschi & 57.2         & Not Found        &9.76  &  49.51$\pm$0.04 \\
               &               &                                           &        & 20.11.2000                 & Laura Maraschi & 58.1         & 16.0             &4.04  &  40.37$\pm$0.04 \\
               &               &                                           &        & 30.11.2001                 & Arvind Parmar  & 44.5         & Not Found        &7.60  &  80.88$\pm$0.05 \\
               &               &                                           &        & 24.05.2002                 & Fred Jansen    & 31.7         & 24.2             &6.54  &  39.51$\pm$0.06 \\
               &               &                                           &        & 24.05.2002                 & Fred Jansen    & 31.6         & Weak QPO?$^b$    &2.90  &  28.17$\pm$0.06 \\
               &               &                                           &        & 24.05.2002                 & Fred Jansen    & 29.8         & Not Found        &20.74 &  38.30$\pm$0.06 \\
               &               &                                           &        & 29.11.2002                 & Arvind Parmar  & 56.7         & 26.8             &10.94 &  21.29$\pm$0.04 \\
               &               &                                           &        & 23.11.2004                 & Fred Jansen    & 39.8         & 15.7             &4.08  &  28.78$\pm$0.05 \\
               &               &                                           &        & 30.11.2005                 & Fred Jansen    & 49.8         & Not Found        &9.34  &  54.52$\pm$0.04 \\
               &               &                                           &        & 07.11.2006                 & Arvind Parmar  & 28.8         & Not Found        &3.94  &  29.68$\pm$0.06 \\
               &               &                                           &        & 22.04.2007                 & Arvind Parmar  & 48.0         & 41.4             &9.93  &  50.69$\pm$0.05 \\
               &               &                                           &        & 12.05.2008                 & Arvind Parmar  & 60.6         & Not Found        &7.32  &  62.26$\pm$0.04 \\\hline
\end{tabular}

$^a$ GTI (Good Time Interval) is determined by screening data for soft-proton flare
using the selection criteria (\# XMMEA$\_{EP}$ and (10000 $<$ PI $<$12000) and
(PATTERN $=$0) for the EPIC/pn camera, and thus requires the total count rate to be
below 0.4 cts/s within the 10--12 keV energy band.

$^b$ See text for details

\end{table*}

\subsection{Data Selection Criteria}

We selected our four target high energy peaked blazars, namely: PKS 0548$-$322, ON 231, 1ES 1426$+$428, 
and PKS 2155$-$304 because they were observed by the European Photon Imaging Camera (EPIC) on board the 
XMM-Newton satellite (Jansen et al.\ 2001).   
Instead of analyzing data from previous or other  current X-ray missions, 
we used data from  XMM-Newton/EPIC due to its excellent detector sensitivity, a
wide field of view that allows for good background subtraction, and the ability to track the source 
continuously for many hours.

The EPIC is composed of three co-aligned X-ray telescopes (Jansen et al.\ 2001) which 
simultaneously observe a source by  accumulating photons in three CCD-based instruments: the twins MOS 1 
and MOS 2 and the pn (Turner et al.\ 2001; Str\"{u}der et al.\ 2001).  The EPIC instrument provides 
imaging and spectroscopy in the energy range from 0.15 to 15 keV with a good angular resolution 
(PSF = 6 arcsec FWHM) and a moderate spectral resolution ($E/\Delta E \approx 20-50$). 

From the XMM-Newton Science Archive\footnote{http://xmm.esac.esa.int/xsa/} 
we downloaded 24 data sets of these four blazars which satisfy the key selection criteria that 
the observation exceeds 7 hours and the source was bright enough to give adequate counts
 in short time bins.  These criteria allow us  to possibly 
obtain reasonable indications of intra-day timescales as well as QPOs with periods up to at least one hour.
These constraints yielded 1 data set for PKS 0548$-$322, 1 for ON 231, 7 observations of 1ES 1416$+$428 
and 15 stares at PKS 2155$-$304.  These 24 observations took place between 30 May 2000 and 
12 May 2008 and lasted between 7.67 hours and  18.97 hours.  In this paper we omit further 
consideration of the observations made on  1 May 2006 of PKS 2155$-$304 which showed a significant QPO 
at $\sim 4.6$ hours and was reported in Lachowicz et al.\ (2009).
The observing log for the remaining 23 X-ray observations is given in Table 1, the first four 
columns of which contain the
blazar name, date of the observation, name of the principal investigator, and length of the data 
for that observation. 

\subsection{Data reduction}

To perform the data reduction we used pipeline products and applied the XMM-Newton Science 
Analysis System (SAS) version 8.0.0 for the LC extraction.  The first step in this reduction 
is to check for strong proton flares in the 10--12 keV range.  We confined our analysis 
to EPIC/pn data as only they were free of soft-proton flaring events and pile-up effects. We 
had to restrict our analysis to the 0.3 -- 10 keV energy range,  as data below 0.3 keV are mostly 
unrelated to bona fide X-rays, and data above 10 keV are usually dominated by background.  Next, 
the data are extracted using a circle of 45 arcsec 
radius centered on the source. Background photons are read out from the same size area located 
about 180 arcsec off the source on the same chipset. Then raw light curves for both the target 
and background areas are  produced. As the final product of data reduction, we obtained source 
LCs for the 0.3--10 keV
band (corrected for background flux and given in unit of counts s$^{-1}$), sampled evenly
with a fixed bin size of $\Delta t = 0.1$ ks. 

In column 5 of Table 1 we report the variability timescale, if present.  The sixth column 
contains the fractional rms variability, defined as 
$\sigma_{frac} = (\sigma^{2}/\mu^{2})^{1/2}$, where $\sigma^{2}$ is the mean variance 
(counts$^{2}$ s$^{-2}$) (Uttley \& McHardy 2001) and $\mu$ is
the mean flux (0.3 $-$ 10 keV, counts s$^{-1}$), which is given in the last column.  
The left columns of panels in Figures 1--4 display these LCs. 

\section{Analysis Techniques}

\subsection{Structure Function Analysis}

We used the first order structure function (SF) for our first time series analysis.
The SF is a powerful tool to quantitatively determine  periodicities and time scales in 
time series data (e.g., Rutman 1978; Simonetti et al.\ 1985; Paltani et al.\ 1997). 
It provides information on the time structure of a data train and it is able to discern 
the range of the characteristic time scales that contribute to the fluctuations. 
It is less affected by any data gaps in the LCs 
and is free of the constant offset in the time series. The first order SF for a data set $a$, 
having uniformly sampled points, is defined as
\begin{eqnarray}
D_{a}(k) = \frac{1}{N_{a}^{'}(k)} \sum_{i=1}^n \omega(i) \omega(i+k)[a(i+k)-a(i)]^{2},
\end{eqnarray}
where $k$ is the time lag, $N_{a}^{'}(k) = \sum \omega(i) \omega(i+k)$, and the weighting 
factor  $\omega(i)$ is 1 if a measurement exists for the $i^{th}$ interval and 0, otherwise.

The square of the uncertainty in the estimated SF is
\begin{eqnarray}
\sigma^{2}(k) = \frac{8 \sigma_{\delta f}^{2}}{N^{'}(k)}D_{a}(k),
\end{eqnarray}
where $\sigma_{\delta f}^{2}$ is the measured noise variance.
Here the lags were examined in units of 0.1 ks, the fixed bin size of our uniformly sampled data.

The behavior of the first order SF indicates the presence (or not) of a timescale in the
LC as follows:
%\begin{enumerate}
%{\item 
a) if the source's first order SF rises and does not display any plateau, it implies that any  time 
scale of its variability is longer than the length of the observation;
b) if there are one or more plateaus, each one may indicate a time scale of variability;
c) if that plateau is followed by a dip in the SF, the temporal lag corresponding to the minimum of 
that dip suggests a possible periodic cycle.
If a LC contains cycles of period $P$, $SF(dt)$ will rise to a maximum at $dt = P/2$ and 
then fall to a minimum at $dt = P$ (Smith et al.\ 1993). When $dt$ is extended to values that are
nearly multiples of $P$, $SF(dt)$ itself goes through a series of cycles.  
Any possible periods extracted from a SF analysis are obtained by averaging the trough-to-trough 
times.  The plots of these SF analyses are presented in the
center columns of Figs.\ 1--4 and the strongest timescales estimated from them are given 
in the fifth column of Table 1.

\begin{table}
{\bf Table 2. The best power-law fits to power spectra$^a$}
\scriptsize

\noindent
\hspace*{-0.2in}
\begin{tabular}{lccrrr} \hline \hline
Source         &  Date of Observation   &  $\alpha$        & log(N) & $\chi^2/\nu$ &$\nu^c$   \\
               &  dd.mm.yyyy            &                  &              &               &         \\\hline
PKS0548$-$322  & 11.10.2002            & $-2.11\pm 0.88$  & $-$9.72  &0.31        &4    \\
ON 231         & 26.06.2002            & $-2.11\pm 0.32$  & $-$5.96  &36.98       &22  \\
1ES1426$+$428  & 16.06.2001            & $-2.44\pm 0.86$  & $-$10.31 &1.19        &4   \\
               & 04.08.2004            & $-2.27\pm 0.86$  & $-$10.32 &0.13        &4   \\
               & 06.08.2004            & $-0.97\pm 0.59$  & $-$4.38  &0.52        &8    \\
               & 24.01.2005            & $-1.00\pm 0.36$  & $-$4.19  &0.15        &18   \\
               & 19.06.2005            & $-2.33\pm 0.76$  & $-$9.79  &0.61        &5    \\
               & 25.06.2005            & $-1.91\pm 1.61$  & $-$8.53  &0.52        &1   \\
               & 04.08.2005            & $-2.46\pm 1.23$  & $-$10.91 &0.20        &2    \\
PKS2155$-$304  & 30.05.2000            & $-3.52\pm 0.76$  & $-$14.68 &0.62        &5    \\
               & 31.05.2000            & $-2.15\pm 0.33$  & $-$8.62  &0.55        &21   \\
               & 19.11.2000            & $-1.91\pm 0.19$  & $-$6.53  &1.21        &55   \\
               & 20.11.2000            & $-2.01\pm 0.29$  & $-$7.71  &5.34        &27    \\
               & 30.11.2001            & $-1.76\pm 0.22$  & $-$6.09  &1.00        &42   \\
               & 24.05.2002            & $-1.88\pm 0.27$  & $-$6.84  &3.13        &29   \\
               & 24.05.2002            & $-1.68\pm 0.38$  & $-$6.32  &0.87        &17    \\
               & 24.05.2002            & $-1.98\pm 0.19$  & $-$6.02  &5.00        &57   \\
               & 29.11.2002            & $-2.65\pm 0.46$  & $-$10.03 &15.63       &12   \\
               & 23.11.2004            & $-1.88\pm 0.37$  & $-$6.88  &2.25        &17   \\
               & 30.11.2005            & $-2.29\pm 0.24$  & $-$8.02  &5.68        &37   \\
               & 07.11.2006            & $-1.91\pm 0.33$  & $-$6.84  &1.47        &21   \\
               & 22.04.2007            & $-2.19\pm 0.24$  & $-$7.79  &1.60        &36   \\
               & 12.05.2008            & $-2.04\pm 0.21$  & $-$7.40  &0.71        &46   \\
\hline
\end{tabular}

$^a$ The power-law model is assumed to be $P(f)\propto f^{\alpha}$ for $\alpha <$ 0 \\
$^b$ The errors are $1\sigma$\\
$^c$ Degrees of freedom (d.o.f.)\\
\end{table}

\subsection{Power Spectral Density}

A classical tool in the search for the nature of temporal variations,
including any periodic and quasi-periodic variability in
a LC, employs calculations of the Fourier power spectral density
(PSD). In general, it is expected that a power-law model, $P(f)\propto
f^{\alpha}$ where $\alpha$ denotes the spectral slope, is a good representation of
the red-noise variability ($\alpha<0$) that characterizes most blazars
(e.g. Kataoka et al.\ 2001; Zhang et al.\ 2002). In general, one assumes
that any significant positive deviation above that red-noise model (i.e., $\gtrsim 3\sigma$) 
can stand for a clear
indication of a quasi-periodic oscillation (QPO) being present in the X-ray
emission (see Vaughan et al.\ 2003a; Vaughan \& Uttley 2006, and references
therein). Because of a lack of high quality data and the apparently low duty cycles 
of their QPOs, the PSDs of powerful AGN were
not  found to display convincing QPO features in studies made over the last three decades.
The very  recent discoveries of QPOs in a narrow line Seyfert 1 galaxy (Gierli{\'n}ski
et al.\ 2008) a FSRQ (Espaillat et al.\ 2008) and a BL Lac (Lachowicz et al.\ 2009) show that 
they are sometimes present and definitely worth searching for.

Here, we employ a standard approach (e.g., van der Klis 1989; Vaughan et al.\
2003a) in which we use the normalization of PSD after Miyamoto et al.\ (1992)
that provides the PSD in units of (rms/mean)$^2$/Hz. We compute all our
Fourier power spectra using the DFT subroutine in MATLAB$^{\copyright}$, and fit
the resultant red-noise part of the PSD, where applicable, using a code based
on Vaughan's (2005) recipe. We model each red-noise component as a single
power-law function,
\begin{equation}
 P(f) = N f^{\alpha} ,
\end{equation}
in the low-frequency portion of the power spectrum, i.e., where the red-noise variability 
at the longest
timescales starts to dominate over the white-noise level present at the highest frequencies.

The results are presented in Figs.\ 1--4, where  the histogram, solid
line, dotted line and horizontal dashed line  denote the PSD, red-noise
component of the PSD, 99.73\% ($3\sigma$) confidence level for the red-noise
model, and the expected Poisson noise level, respectively. Table 2 lists the
best values for the two parameters of the fitted red-noise model (a slope $\alpha$ and 
normalization $N$) and the  goodness-of-fit measure, $\chi^2/\nu$, for the number of
degrees-of-freedom, $\nu = m-2$, where $m$ equals the number of frequency bins involved
in the fit.

\section{Results}   

\subsection{PKS 0548$-$322}

PKS 0548$-$322 
is a nearby BL Lac object and is hosted in a giant elliptical galaxy 
(Falomo Pesce \& Treves 1995; Wurtz et al.\ 1996) which is the dominant member of a rich 
cluster of galaxies. Imaging and spectroscopic observations of the field 
around the source show that it is at a redshift of $z = 0.069$ (Falomo et al.\  1995).
The synchrotron power peaks in the X-ray band, and for this reason it is classified as an 
HBL source (Padovani \& Giommi 1995). 

In the span of 30 nights of observations, (Cruz-Gonzalez \& Huchra 1984) noticed 
a typical $\sim$0.02 mag (2\%) day$^{-1}$ optical variability in the source.  The source was 
monitored by Xie et al.\ (1996) in the V-band for IDV and they once reported a rise of 0.58 magnitudes over an interval of 5 minutes.
Optical IDV of this source was also studied for a few nights in 1995--1996 by Bai et al.\ (1998).

PKS 0548$-$322 is a relatively strong and rapidly variable source in the X-ray energy 
bands (Blustin et al.\ 2004). 
The possibility of deviation from a single power law in the X-ray spectrum has been present since 
early observations of this source. It was observed by EXOSAT during 1983 to 1986 (Barr et al.\ 1988; 
Garilli \& Maccagni 1990) and its SED peaked in the  2.5--5 keV energy range. 
GINGA observations in February, 1991 (Tashiro et al.\ 1995) of 
the 2--30 keV energy range  showed a flatter spectrum, with the synchrotron peak of the SED moved to 
energies higher than 30 keV. 
This blazar was observed by BeppoSAX in 1999 during February through April  (Costamante 
et al.\ 2001). 
PKS 0548$-$322 was observed by 
the Swift satellite  (Gehrels et al.\ 2004) during April through June of 2005 and the spectral analysis 
of its  X-ray telescope (XRT) data was reported by Burrows et al.\ (2005) and that from the 
Ultraviolet/Optical telescope (UVOT) was discussed  by 
Roming et al.\ (2005). These data confirmed that the X-ray spectrum of this HBL shows a well established 
curvature (Perri et al.\ 2007) for which evidence had been found by Costamante et al.\ (2001).

Stecker et al.\ (1996)  predicted VHE gamma-ray fluxes from this object. 
The CANGAROO 
(Collaboration of Australia and Nippon  for a  GAmma Ray Observatory in the Outback) group have reported 
limits to the VHE gamma-ray emission above $\sim$ 1.5 TeV for the source (Roberts et al.\ 1998). VHE 
gamma-ray observations were also made with the Mark 6 telescope (Armstrong et al.\ 1999) and 3$\sigma$ 
limits to the VHE gamma-ray flux of 2.4 $\times$ 10$^{-11}$ cm$^{-2}$ sec$^{-1}$ above 300 GeV for the 
source was reported by Chadwick et al.\ (2000). The source was, however, detected at the threshold energy 
of 190 GeV by HESS with an integrated flux of 6.65$\times$10$^{-12}$ cm$^{-2}$ s$^{-1}$ (Aharonian et al. 
2005a).  This source has not yet been detected by the FERMI satellite (Abdo et al. 2010).

This blazar PKS 0548$-$322 was continuously monitored by XMM-Newton for 45.5 ks on October 11, 
2002 and the rather flat LC we extracted from this data is given in the first panel 
in Fig.\ 1.  The SF we produced  from this observation is plotted as the middle panel in the first 
row of Fig.\ 1.
This SF plot shows a continuous rising trend as the temporal lag increases.  Therefore any variability 
timescale that might have been present at that time  is larger than the length of the data set.

The power spectrum analysis of our single data set for  PKS 0548$-$322 revealed the variability pattern 
to be in a good
agreement with a model  dominated by  a broad-band white noise process, i.e.,
$P(f)\propto f^{0}$.  However, at the longest accessible timescales (or lowest frequencies)
there is good evidence for a contribution from red-noise variability  (right column of the first row 
in Fig.\ 1 and Table 2).

\subsection{ON 231}

ON 231 was  previously known as the ``variable star'' W Comae (and is still also called by that name). 
In 1971, because of its odd properties, particularly a  strong optical variability and an 
apparently line-less continuum, 
it was suggested that it was an extragalactic source and was assigned to the then rather new class 
of BL Lac objects (Biraud 1971; Browne 1971). The later detection of weak emission lines in 
its spectrum during a time when the synchrotron optical emission was less dominant
made it possible to determine its redshift, $z =0.102$ (Weistrop et al.\ 1985). 

The historical B passband LC of the source (during 1935--1997) was plotted by Tosti et al.\ (1998). It 
has shown optical flux variations on diverse timescales ranging from a few hours to several years 
(Xie et al.\ 1992; Smith \& Nair 1995).
In 1995, ON 231 started a phase of strong brightness and activity that culminated in an
outburst in April--May 1998, when it reached its brightest magnitude at R = 12.2 (Massaro et 
al.\ 1999) comparable to that measured at the beginning of the past century when it was discovered as a 
variable star (Wolf 1916). 
After the optical outburst, ON 231  showed 
a slow decline in its mean luminosity (Tosti et al.\ 2002). 
Photometric observations on January  2007 in the Johnson R passband found  the source
to be in a low state (Gupta et al.\ 2008) comparable to the faintest state ever noted by Tosti 
et al.\ (1998). 

BeppoSAX observed the source in May 1998 with good sensitivity and spectral resolution, following 
the exceptional 
outburst that occurred in April--May of that year.
The sensitivity of the WFC (Wide Field Camera) on BeppoSAX depends on the pointing direction,
but for an AGN at  high-galactic latitude it is possible to determine the spectrum up to 200 keV 
for sources with fluxes down to about 1mCrab (Boella et al.\ 1997).
An X-ray spectrum was measured 
from 0.1--100 keV and a smaller variability amplitude between 0.4--10 keV was detected 
(Tagliaferri et al.\ 2000). Massaro et al.\ (2008)  studied the spectral energy distribution 
of the source using Swift UV data, along with soft X-ray data from three Swift/XRT pointings. 

ON 231 was discovered as a gamma-ray source (von Montigny et al.\ 1995) 
and was detected by EGRET in the 100 MeV to 10 GeV band
(Hartman et al.\ 1999).  This source exhibited the hardest spectrum among the EGRET AGN detections 
(Sreekumar et al.\ 1996). 
A strong gamma$-$ray outburst in W Com which lasted for 4 days was observed in March 2008; 
VHE gamma-ray emission was detected by VERITAS with a statistical significance of 4.9 $\sigma$ 
for the data set of January--April 2008 (Acciari et al.\ 2008).   

The XMM-Newton satellite  continuously monitored ON~231 for 27.6 ks on 26 June 2002 and its LC and SF are 
plotted in the second row of Fig.\ 1.  The LC shows considerable variability, with the flux changing 
by a factor of about three during this, the shortest of all the observations we considered. 
The SF plot starts with a continuous rising trend  to a maximum, followed by a dip;  the SF then rises and dips  two more times.  The peaks in the SF are at 10.4, 15.7 and 22.6 ks, while the minima following them are
at 12.0, 19.9 and 27.1 ks, respectively.  The last of these dips is too close to the total length of the 
observation to be meaningful.  The nominal 
variability time scales are taken as the differences between maxima and previous minima,
or 10.4 ks, 3.7 ks ($= 15.7-12.0$) and  2.7 ks ($= 22.6-19.9$).  The first dip might provide evidence 
for a periodicity at 12.0 ks, but as the subsequent dips are not at multiples of this value, no claim 
for periodicity can be made from this LC using the SF.

Our Fourier analysis of the LC of ON~231 is given in the last column of the second row of Fig.\ 1.  
It reveals that the overall profile of the 
broad-band variability of ON~231 can be  described by a single power-law with
a slope $\alpha=-2.11\pm 0.32$, extending down to timescales of $1/f\simeq
1000$~s. (This best fit is achieved, however, at a poor reduced $\chi^2/\nu=36.98$ for
$\nu=22$ degrees of freedom.)  This, and other (usually much better) best fits to the PSD slopes for 
the other blazars, along with
the number of degrees-of-freedom  used to estimate the red-noise portion of the PSD, are given in Table 2.
Interestingly, this PSD also showed a statistically insignificant ($<2 \sigma$) peak
located at $\sim  1.7\times 10^{-4}$~Hz, suggesting a possible very weak QPO
with period close to $\sim \! 5.9$~ks. However, as this possible timescale differs 
from any of  those hinted at by the SF for ON~231, we certainly cannot claim that any significant 
quasi-periodicity was present in this blazar at the time of this observation.

\subsection{1ES 1426$+$428}

The blazar 1ES 1426$+$428 
now thought to be at  $z=0.129$ was discovered in the medium X-ray band (2--6 keV) with the Large 
Area  Sky Survey experiment (LASS) on HEAO-1 (Wood et al.\ 1984). It was 
classified as a BL Lac object on the basis of its featureless optical spectrum. Observations made with the 
MDM 2.4 m Hiltner telescope constrained its redshift to z $\geq$ 0.106 (Finke et al.\ 2008) and its host galaxy 
was resolved by Urry et al.\ (2000). Remillard et al.\ (1989) reported its radio flux is 33.7 mJy at 1.4 GHz 
and they also presented its optical spectrum. 

In X-rays, 1ES 1426$+$428  is bright, with a 2--6 keV luminosity $\sim$ 10$^{44}$ ergs sec$^{-1}$
(assuming isotropic emission), which is typical of HEAO-1 (High Energy Astronomical Observatory) BL Lac objects 
(Schwartz et al.\ 1989). X-ray observations of the source were carried out by Costamante et al.\ (2001) 
with BeppoSAX and 
the X-ray spectrum was found to extend up to 100 keV. An X-ray absorption feature was detected in 
1ES 1426+428 by 
Sambruna et al.\ (1997) which led to the exploration of the gaseous environment of  BL Lac objects. 
Falcone et al.\ (2004) 
observed 1ES 1426+428 with the Rossi X-Ray Timing Explorer (RXTE) and found that the peak energy in 
the SED was sometimes 
in the excess of  100 keV and at other times in the 2.4--24 keV region. 

This source has been observed in the TeV gamma-ray band by CAT (Cerenkov Array at Themis) and HEGRA 
(High Energy Gamma Ray Astronomy) (Djannati-Atai et al.\ 2002; Petry et al.\ 2000; Aharonian et al.\ 2002, 
2003; Horan et al.\ 2002). Observations made with the MAGIC telescope gave a upper limit of $\gamma-$ray 
flux at 200 GeV is 5.5 $\times$ 10$^{-12}$ ergs cm$^{-2}$ s$^{-1}$ (Albert et al.\ 2008). The intrinsic 
spectrum of 1ES 1426$+$428 was inferred by Costamante et al.\ (2003) using a model spectrum of the 
extragalactic background light (EBL) given by Primack et al.\ (2001). Aharonian et al.\ (2003) 
obtained a rising intrinsic source spectrum in the TeV regime, using the same EBL spectra 
that Costamante et al.\ (2003) and Malkan \& Stecker (2001) adopted.

The blazar 1ES 1426$+$428  was continuously monitored by XMM-Newton for 62.0 ks on 16 June  2001 and 
its LC and SF on that day  are plotted in the third row of Fig.\ 1. The SF plot shows a continuous 
rising trend followed by a dip which indicates variability time scale of 46.8 ks. However, this was the 
only possible time scale seen for this blazar, which was also continuously monitored for 61.9, 68.3, 
29.8, 37.9, 33.9, 39.9 ks on 4 August 2004, 6 August 2004, 24 January 
2005, 19 June 2005, 25 June 2005, and 4 August 2005, respectively. All these LCs and their SFs are plotted in  
Figs.\ 1 and 2. The SF plots of all these LCs have continuous rising trends, so any variability 
timescales associated with these 
LCs are longer than the lengths of the observations.

We have found that the Fourier power spectra for four of the seven observations of 1ES
1426$+$428 blazar do not display any significant red-noise components, with only
possible indications of red-noise at the lowest few bins in temporal frequency. Power-law
red-noise is best seen on 16 June 2001 ($\alpha=-2.44\pm 0.86$;
$\chi^2/\nu=1.19$ for $\nu=4$). Red noise is  weakly detectable
on 6 August 2004 and 24 January 2005, respectively (see the 
right columns of Figs.\ 1 and 2 and Table 2). 
Neither the SF nor PSD techniques indicate any hints of QPOs in this blazar, and only
one indication of a timescale of variability was detected via the SF analysis at 46.8 ks,
which was longer than four of the seven observations.

\subsection{PKS 2155$-$304}

PKS 2155$-$304  was first identified as a BL Lac by Schwartz et al.\ (1979) and Hewitt \& Burbidge (1980), 
and is the brightest BL Lac object from the  UV 
to TeV  energies in the southern hemisphere. It had already been  observed in the radio as part
of the Parkes survey (Shimmins \& Bolton 1974). Since the 1970's, it has been observed on diverse 
timescales by many space and ground based telescopes in all  EM bands 
(e.g., Carini \& Miller 1992; Urry et al.\ 1993; Brinkmann et al.\ 1994; Marshall et al.\ 2001;
Aharonian et al.\ 2005b; Dominici, Abraham \& Galo 2006; Dolcini et al.\ 2007; Piner, Pant \&
Edwards 2008; Sakamoto et al.\ 2008; and references therein). A redshift of 0.117$\pm$0.002 was 
given by Bowyer et al.\ (1984) but  careful spectroscopy of galaxies in the field of PKS 2155$-$304, 
by Falomo et al.\ (1993) demonstrated that the 
redshift of the source is 0.116$\pm$0.002. Simultaneous observations of PKS 2155$-$304 in the optical and 
near-infrared bands showed significant variability on a wide range of timescales (e.g., 
Domini, Abraham \& Galo 2006; Osterman et al.\ 2007; Dolcini et al.\ 2007).  PKS 2155$-$304 has 
shown a linear optical polarization of 3$-$7\% with a polarization  position angle varying 
between 70$^{\circ}$ and 120$^{\circ}$ (Tommasi et al.\ 2000).

Schwartz et al.\ (1979) reported the first X-ray observations of PKS 2155$-$304  using HEAO-1. 
It has been suggested that the synchrotron emission of PKS 2155$-$304 tends to peak in the UV-EUV rather than
in the X-rays (Zhang 2008). Simultaneous multi-wavelength observations of this blazar from 
X-ray  (using XMM-Newton) and optical bands are presented by Zhang et al.\ (2006a). Using XMM-Newton  
observations from EPIC pn data, hardness ratio variations and temporal cross-correlations between bands 
were computed and discussed by Zhang et al.\ (2006b) who found the variations between bands were more 
strongly correlated during flares.

PKS 2155$-$304  was classified as a TeV blazar by the detection of VHE gamma-rays by the Durham MK 6 
telescopes (Chadwick et al.\ 1999). It was confirmed as a high energy gamma-ray emitter by HESS at a 
45 $\sigma$ significance level (Aharonian et al.\ 2005b) and it is so bright that IDV is also detectable 
in the this VHE band. 

Observations in the UV by the International Ultraviolet Explorer of PKS 2155$-$304 seemed to show 
a short-lived quasi-period
of $\sim$0.7 day, but only a few cycles were present, so it was not a firm detection (Urry et al.\ 1993). 
Recently, we have shown nominally very strong evidence for  a $\sim$4.6 hour QPO in this source in  
XMM-Newton observations made on 1 May 2006; but again, as only $\sim$4 cycles of this fluctuation were 
seen during about 18 hours of the observation,  the QPO indication is not iron-clad (Lachowicz et al.\ 2009).

The blazar PKS 2155$-$304  was continuously monitored by XMM-Newton for 57.2, 44.5, 29.8, 49.8, 28.8, 
and 60.6 ks on 19 November 2000, 30 November 2001, 24 May 2002, 30 November 2005, 07 November 2006, and 
12 May 2008, respectively. There was no evidence for timescales on those days. These LCs and their SFs are 
plotted in Figs.\ 2--4. As the SF plots of all these light curves show continuous rising trends, any 
variability timescale for these LCs are longer than the  lengths of their data trains.

This source also was continuously monitored for 37.9 ks on 30 May 2000, 59.2 ks on 31 May 2000, 58.1 ks on 
20 November 2000, 31.7 ks on 24 May 2002, 56.7 ks on 29 November 2002, 39.8 ks on 23 November 
2004, and 48.0 ks on 22 April 2007 and during these observations indications of timescales were found.  
These LCs and their SFs are also plotted in Figs.\ 1 and 2. Each of these SF plots show a continuous 
rising trend followed by a dip. These dips indicate possible variability time scales 
of 24.2 ks, 41.2 ks, 16.0 ks, 24.2 ks, 26.8 ks, 15.7 ks and 41.4 ks, respectively. 

During the second observation made of  PKS 2155$-$304 on 24 May 2002), we detected four dips in 
the SF that might indicate a weak QPO signal (see the fourth row of Fig.\ 3).  For this data set the
SF remains quite flat for the first $\sim$ 5.3 ks, indicative of white noise domination
at short lags or high frequencies (e.g., Paltani et al.\ 1997).  The SF then rises to a peak at 8.8 ks, 
indicating a  timescale of 3.5 ks, where we follow Paltani et al.\ (1997) and subtract the 5.3 ks before 
the SF started its rise.  The first dip in the SF is at  $\sim$12.2 ks, providing a hint of a period at 
6.9 ks ($= 12.2 -5.3$ ks). The timescales for the subsequent rises to peaks following minima are 2.6 ks, 
2.5 ks and 3.0 ks, while the three subsequent dips yield ``periods'' of 4.8 ks, 4.6 ks and 7.0 ks, as 
measured from the previous dips.  Although these times between dips are clearly not the same, and 
therefore do not give a clear indication of a single period, if one averages the four ``periods'' obtained 
from the dip intervals them one obtains a broad ``quasi-period'' of $5.5 \pm 1.3$ ks, although two separate 
timescales of $\sim 4.7$ ks and $\sim 7.0$ ks are an alternative possibility. For the same observation, the PSD analysis
provides us with a possible indication of a weak ($< 2\sigma$) peak centered around
$\sim\! 1.7\times 10^{-4}$~Hz (see Fig.\ 3). That would also suggest a QPO period of $\sim\!
5.9$~ks, which is, coincidentally, a value close to that  we noted  for a possible QPO in
ON~231.

A quantitative analysis of all 14 PSDs of PKS 2155$-$304 allowed us to find its
broad-band X-ray variability to be very well described for most observations by a single power-law
component denoting red-noise  at lower frequencies with white noise present at high frequencies.  As 
seen from Figs.\ 2--4, for about half of these data sets this power-law extended over the  great 
majority of the frequency band we could measure, while for the other half the white noise dominated 
a significant portion of it. The results of the best fit slope values over the red-noise dominated 
portions are given  in Table 2 and range from $-3.5 \leq \alpha \leq -1.7$.   In the data discussed here
the only  hint of quasi-periodic variability  detected in the X-ray PSDs  of this 
was the observation on 24 May 2002 mentioned in the previous paragraph.
 
\section {Discussion} 

The generally accepted models for the strong variability in blazars on  LTV timescales (months to decades)
involve shocks propagating down relativistic jets pointing close to the direction from which we observe them 
(e.g.\ Scheuer \& Readhead 1979; Marsher \& Gear 1985; Wagner \& Witzel 1995). These shocks moving through 
jets  almost certainly dominate the fluctuations in BL Lacs at essentially all phases and in FSRQs when 
they are in high flux states. The shorter STV and IDV variations can be most simply explained in terms of 
irregularities in the jet flows and any possible periodic variability has been suggested to arise from 
shocks passing through nearly helical structures in a jet (e.g., Camenzind \& Krockenberger 1992), by 
jets whose orientation with respect to us changes (e.g., Gopal-Krishna \& Wiita 1992) or by turbulence 
behind a shock (Marscher,  Gear \& Travis 1992).
Instabilities on or above accretion disks also should play a role in luminosity fluctuations for FSRQs, 
particularly in lower luminosity states, and they presumably dominate in 
non-blazar AGNs such as Seyfert galaxies (e.g., Abramowicz et al.\ 1991; Mangalam \& Wiita 1993; 
Wagner \& Witzel 1995).

The X-ray emission we are concerned with here, however, is unlikely to arise from the disk itself and 
is usually attributed 
to coronal flares above the disk in non-blazar AGN (e.g.\ di Matteo 1998) and to jet emission in
blazars (e.g.\ B{\"o}ttcher 2007). Some recent work has argued that the nuclear emission in low 
luminosity radio-loud AGNs indicates that the accretion disk is radiatively inefficient
(e.g., Balmaverde et al.\ 2006; Balmaverde \& Capetti 2006).  If this is correct, the emission in 
even the low luminosity radio-loud AGNs that are likely the parent population of BL Lacs is dominated 
by non-thermal emission from the base of the jet. 

Now X-ray flares not far above the inner part of a disk might well 
have fluctuations closely  related to the disk's orbital period.  
As most of the quasi-thermal radiation from an accretion disk emerges from its innermost portion, the 
relevant orbital
period and/or size of the emitting region is most likely to be at, or at least close to, that of the last 
stable circular orbit around the central supermassive black hole (SMBH) that is allowed by general 
relativity. If the variability does arise in this way, then one can compute a rough estimate of 
the mass, $M$, of the SMBH   which is proportional to the period (e.g, Gupta et al.\ 2009): 
\begin{equation}
M / M_{\odot} = 3.23 \times 10^4 P(s) [(r^{3/2} + (a/M)]^{-1} (1+z)^{-1}
\end{equation}
where $P(s)$ is 
a period of some fluctuations (in seconds), $r$ is the radius of the emitting region within the disk 
in units 
of $GM/c^2$ and $a/M$ is the dimensionless BH angular momentum parameter.  For the possible periods of 
$\sim 5.9$ ks seen in one PKS 2155$-$304 data set and in the single ON 231 data set this corresponds to 
$\sim 1.2 \times 10^{7}$ M$_{\odot}$ for a non-rotating SMBH (with $r = 6$, the last stable circular 
orbit for a disk around a Schwarzschild BH with $a/M = 0$)
and $\sim 7.4 \times 10^{7}$ M$_{\odot}$ for a rapidly rotating SMBH (with $r = 1.2$, 
the last stable circular orbit for a co-rotating disk around an physically maximally spinning
BH with $a/M = 0.998$; e.g., Thorne 1974).  We again wish to stress that neither of those QPO signals 
are statistically significant, as their peaks in the PSD
plots  do not reach  the usually adopted $3\sigma$ threshold (Vaughan 2005)
for significance of a QPO  with respect  to the underlying red-noise source
variability defined by a single power-law model.

Still, because most blazar emission is expected to arise from jets that are launched from the immediate 
vicinity of the SMBH rather than the disk or its corona, 
(e.g.\ Marscher et al.\ 2008), the observed timescale, $P_{obs}$, of any fluctuation is likely to be
reduced with respect to the rest-frame timescale, $P_{em}$, by the Doppler factor, $\delta$, as well as 
increased by a factor of $(1+z)$. As usual, the Doppler factor, $\delta$,  is defined in terms of  the 
velocity of the jet (and/or the shock propagating down the jet), $V$, and the angle between the jet 
and the observer's line-of-sight, 
$\theta$, as $\delta = [\Gamma (1-\beta~{\rm cos}\theta)]^{-1}$; we have adopted the standard
notation, $\beta = V/c$ and 
$\Gamma = (1-\beta^2)^{-1/2}$.  Hence, $P_{obs} = P_{em}/\delta$.  From very high energy gamma-ray 
observations of PKS 0548$-$322, $18 < \delta < 25$ was estimated (Superina et al.\ 2008). 
Radio
observations indicate that  ON 231 has  $1.2 < \delta < 5.2$  (Wu et al.\ 2007; Hovatta et al.\ 2009).
 From
INTEGRAL observations at 150 keV, $\delta\simeq$ 27.3 was proposed for 1ES 1426$+$428 (Wolter et al.\ 2008).
Finally, from high to very high energy gamma$-$ray observations  $\delta>$30 was obtained for PKS 2155$-$304
(Urry et al.\ 1997; Ghisellini \& Tavecchio 2008).  
The fluctuations could be induced by the disk rotation and then advected into the jet where they could 
become responsible for variations in the jet density, velocity or magnetic field (e.g., Wiita 2006). 
In that case the
amplitude of the variation would be greatly enhanced by Doppler boosting (roughly as $\delta^{3}$)
while time-compressed by $\delta^{-1}$ (e.g., Gopal-Krishna et al.\ 2003). This boosting could possibly 
make even weak disk variations 
detectable once amplified by the jets. In that case the SMBH mass estimates would rise by a factor of $\delta$.   
Taking the average of the $\delta$ values quoted for ON 231 (3.2), the nominal SMBH mass estimates
 would rise to $3.8 \times 10^{7}$ M$_{\odot}$ for the non-rotating case and  
$2.4 \times 10^{8}$ M$_{\odot}$ for the rapidly spinning case.   If $\delta = 30$ for PKS 2155$-$304 
then the estimates for
the SMBH mass  would be between $3.6 \times 10^{8}$ M$_{\odot}$ and
$2.2 \times 10^{9}$ M$_{\odot}$, with the bounds corresponding to non- and rapidly- rotating
situations.
Such higher values  are in better agreement with most estimates for the masses of SMBHs,
($> 2  \times 10^{8}$ M$_{\odot}$) in powerful  radio loud AGN  (e.g., Dunlop et al.\ 2003),
but those masses may not be directly applicable to HBLs.
Because the galaxies surrounding 
the AGN in most HBLs, including ON 231 and PKS 2155$-$304, are barely detected and the optical 
spectra of the AGN are essentially featureless,
no other techniques usually can be applied to provide alternative SMBH mass estimates
for  these HBLs.   However, to give an indicative mass for an HBL, we note that
stellar velocity dispersions have been measured for a 
number of low-$z$ BL Lacs, including 0548$-$322, for which a SMBH mass of $\simeq 1.4 \times 10^8$M$_{\odot}$ 
has been 
estimated (Barth, Ho \& Sargent 2003).

\section {Conclusions}

From our structure function analysis of  high quality archival X-ray light curves of four HBLs taken by 
the XMM-Newton EPIC/pn detector  between May 2000 to May 2008, we found only one apparently 
significant possible detection of a QPO which we have presented in detail elsewhere (Lachowicz et al.\ 2009).  
This was discovered using the SF method for  PKS 2155$-$304 at a timescale of $\approx$16.7 ks. This 
estimated  QPO  was confirmed by three other essentially 
independent methods, viz,  a periodogram analysis, a wavelet analysis and a careful  power
spectrum density analysis along the lines of Vaughan (2005) that gave a
probability of $> 0.9973$ that  a real signal had been seen; however, the short
length of the data train compared to the nominal period still means that we
cannot make a firm claim for the presence of a QPO (Lachowicz et al.\ 2009). 

Here we have presented  the remaining  23  LCs, none of which showed any significant detections of QPOs. 
One out of these 14 observations of PKS 2155$-$304 showed four dips in the SF, which might conceivably 
indicate a a broad QPO around 5.8 ks but a PSD analysis of the same observation gave a 
only a very weak hint of one at 5.9 ks.  
The one data set available for ON 231 showed three dips in the SF that did not provide
any significant period while the PSD analysis  gave a weak signal corresponding to a possible 
QPO around 5.9 ks.  We stress that no significant signal was seen in the PSD plots for any 
of these 23 observations. We did find evidence  for time scales ranging from 15.7 ks to 46.8 ks in 
eight additional LCs using SF analyses; one of these was for 1ES 1426$+$428 and the other seven for 
PKS 2155$-$304.  Combining our new results for PKS 2155$-$304 along with those of
Urry et al.\ (1993) and Lachowicz et al.\ (2009) it is clear that this blazar is well worth continued 
observations to search for timescales and possible QPOs. For the remaining 13 LCs, any variability 
timescale that may be present seems to be longer than the length of the X-ray data sets, which ranged 
from 29 to 68 ks.  For none of these data sets were the individual measurements long enough to clearly 
see a break in the PSD at low frequencies that could also provide a handle on SMBH masses 
(e.g., Marshall, Ryle \& Miller 2007).  Although multi-wavelength 
campaigns are necessary to properly understand the emission regions and processes in blazars, lengthy and 
frequent monitoring in single bands, such as has been done with XMM-Newton, can provide interesting constraints.

\acknowledgements
We thank the anonymous referee for many suggestions that significantly improved the presentation
of this paper.
This research is based on observations obtained with XMM$-$Newton, an ESA science mission with instruments
and contributions directly funded by ESA member states and NASA. 

{}

\clearpage

\begin{figure}
 \centering
  \includegraphics[width=6.0in,height=3.0in]{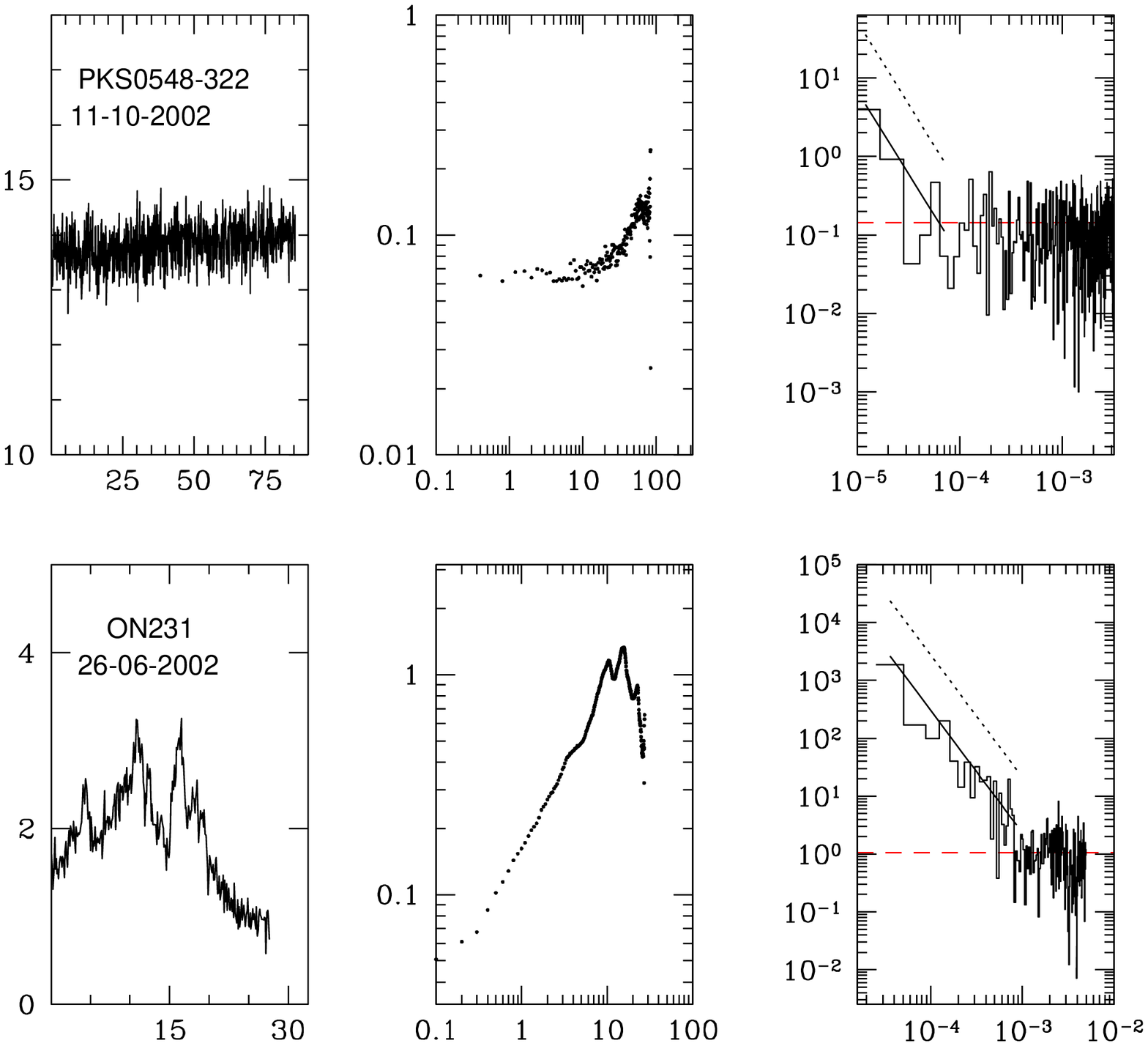}
  \includegraphics[width=6.0in,height=3.0in]{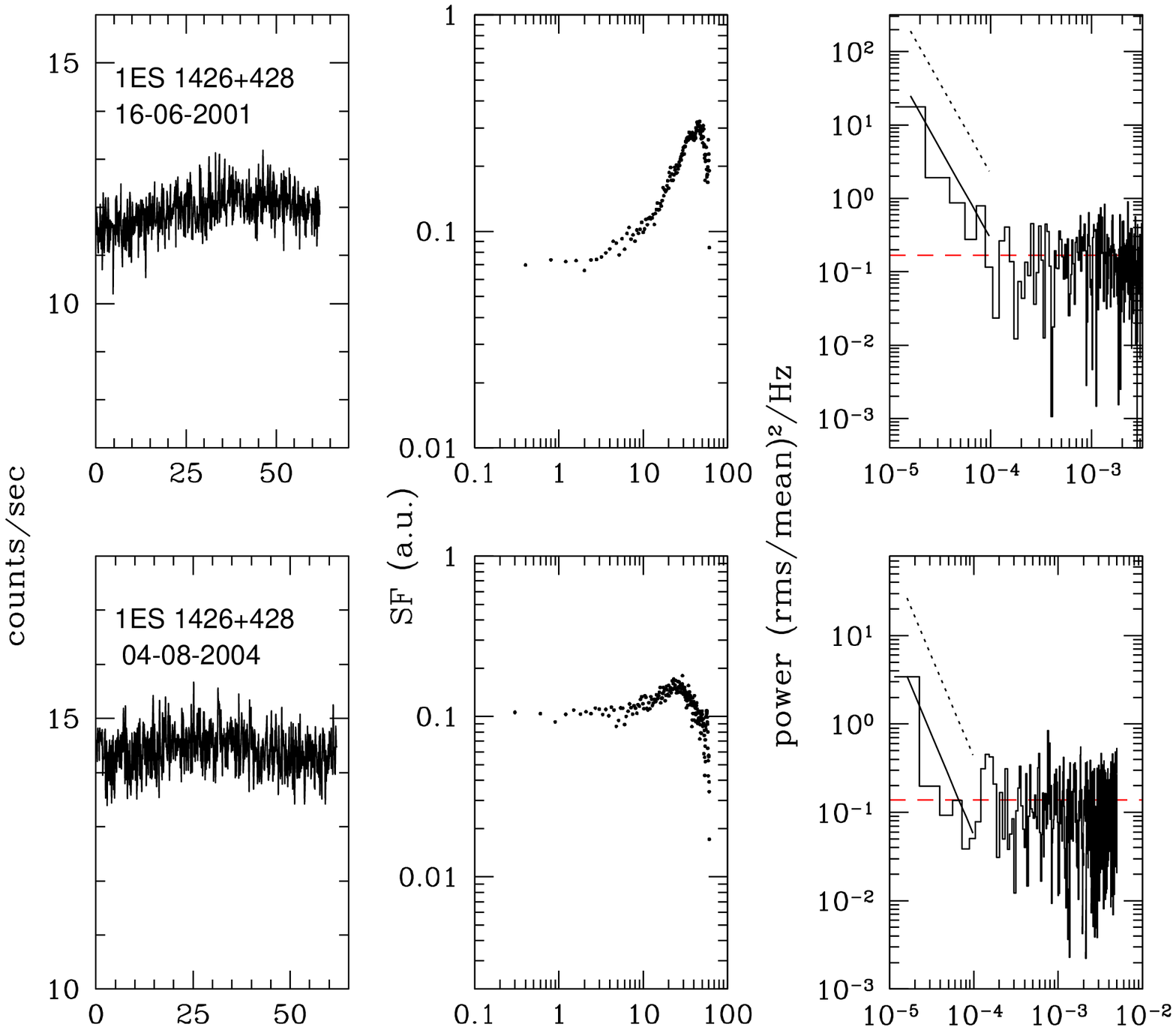}
  \includegraphics[width=6.0in,height=3.0in]{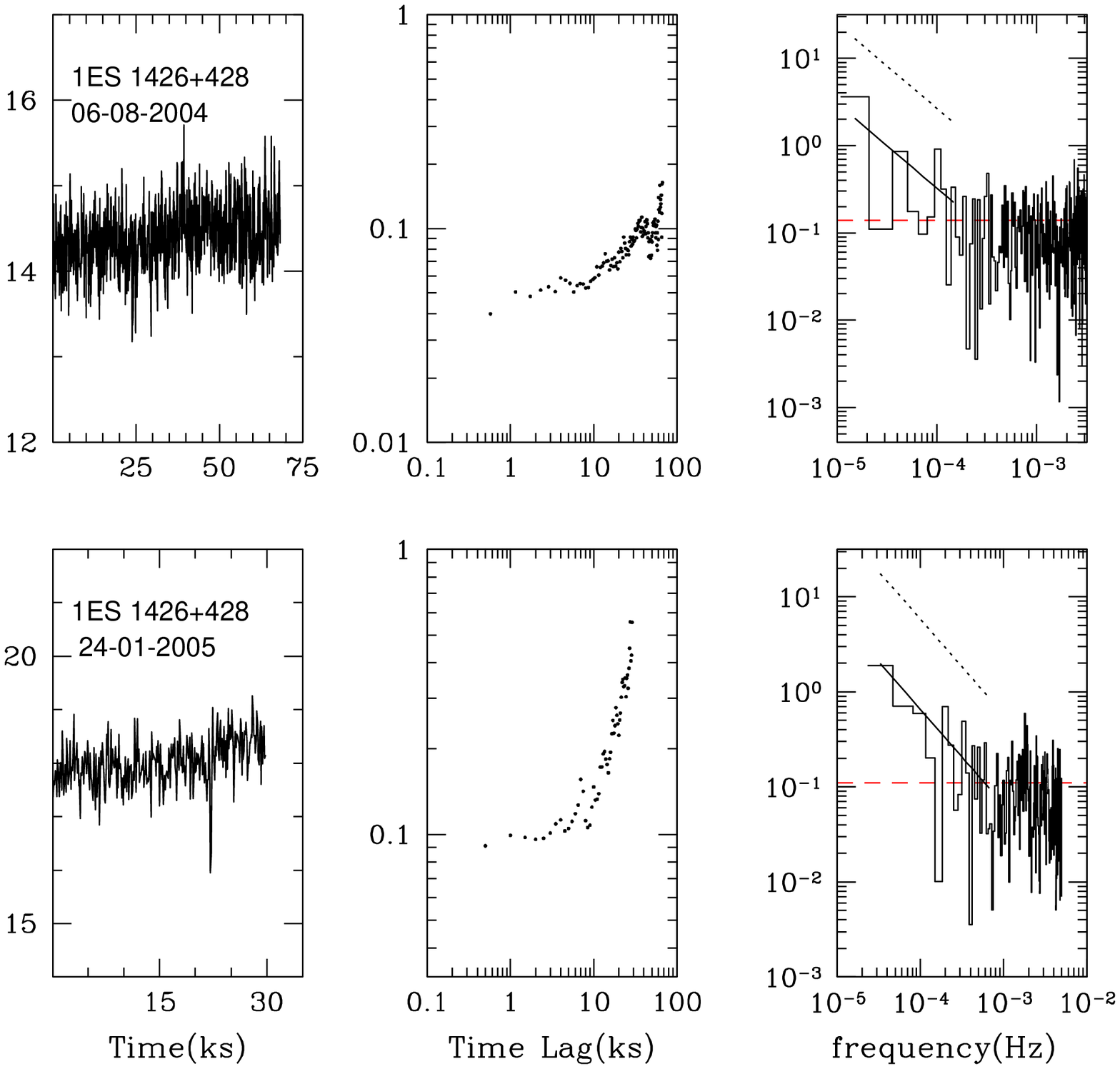}
\caption{Light curves (left panels), structure functions (middle panels) and
power spectral densities (right panels) of PKS0548$-$322, ON 231, and 1ES 1426$+$428.
The source and dates of each observation are given in the light curve panels.}
\end{figure}

\clearpage

\begin{figure}
 \centering
  \includegraphics[width=6.0in,height=3.0in]{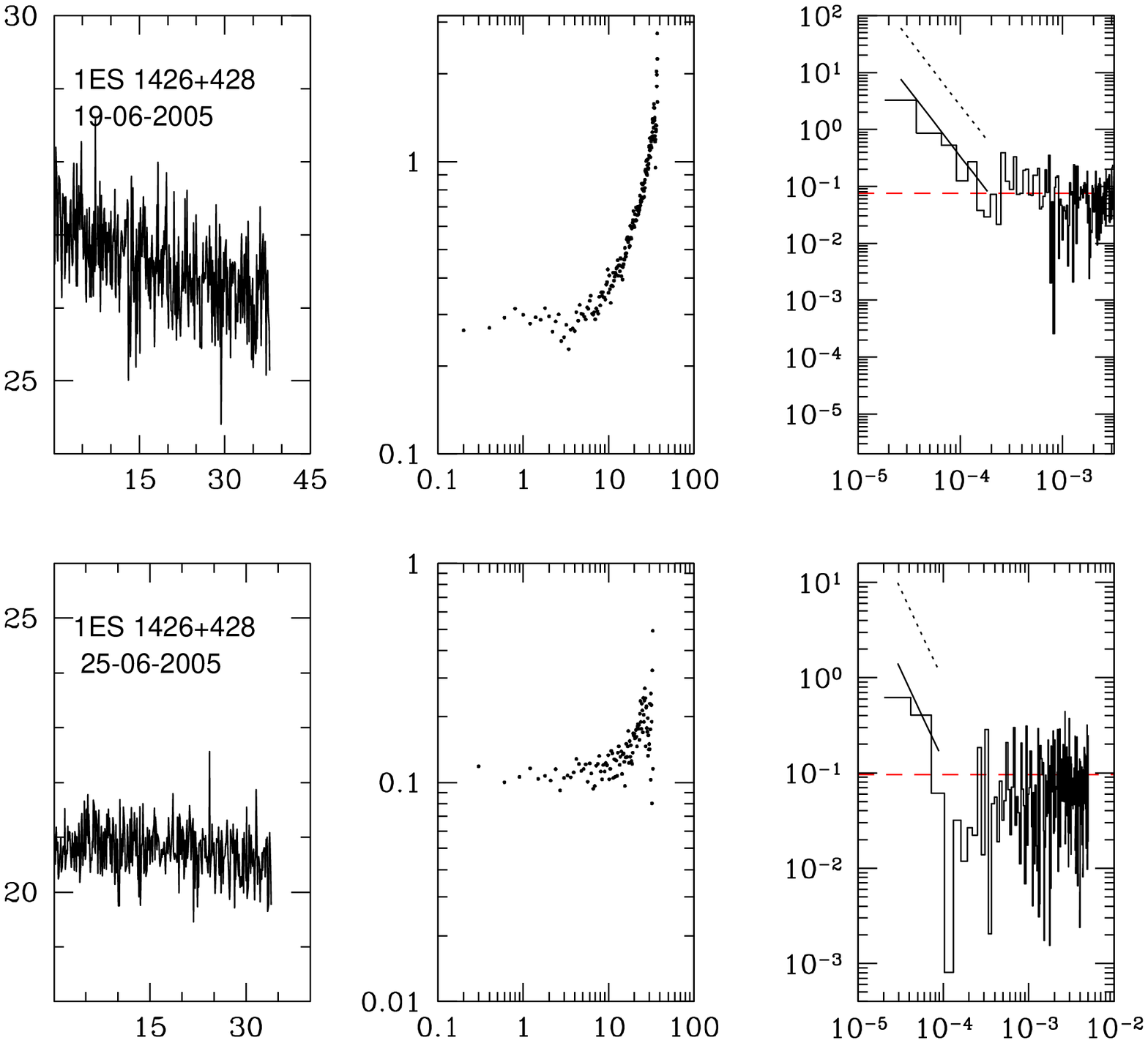}
  \includegraphics[width=6.0in,height=3.0in]{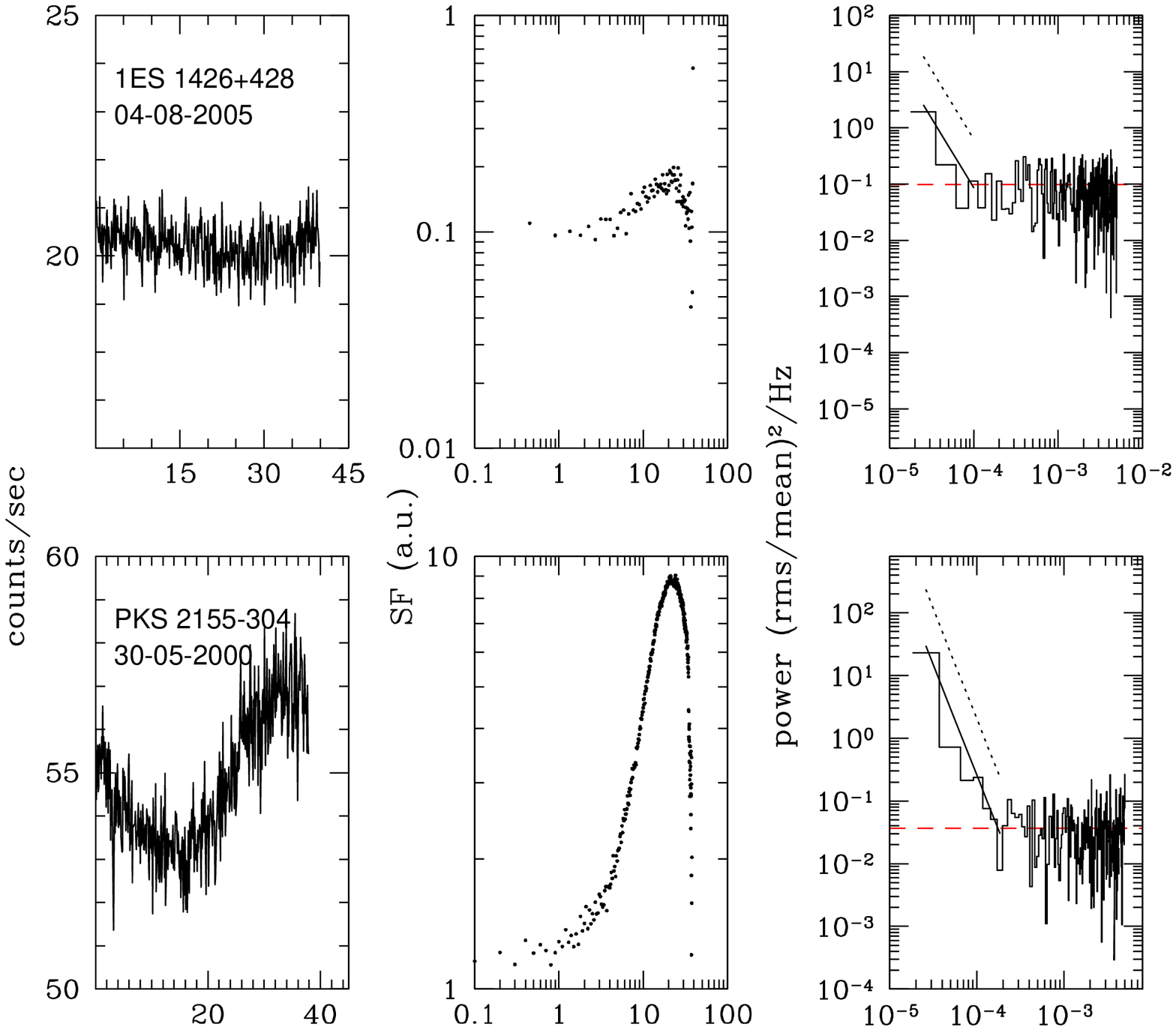}
  \includegraphics[width=6.0in,height=3.0in]{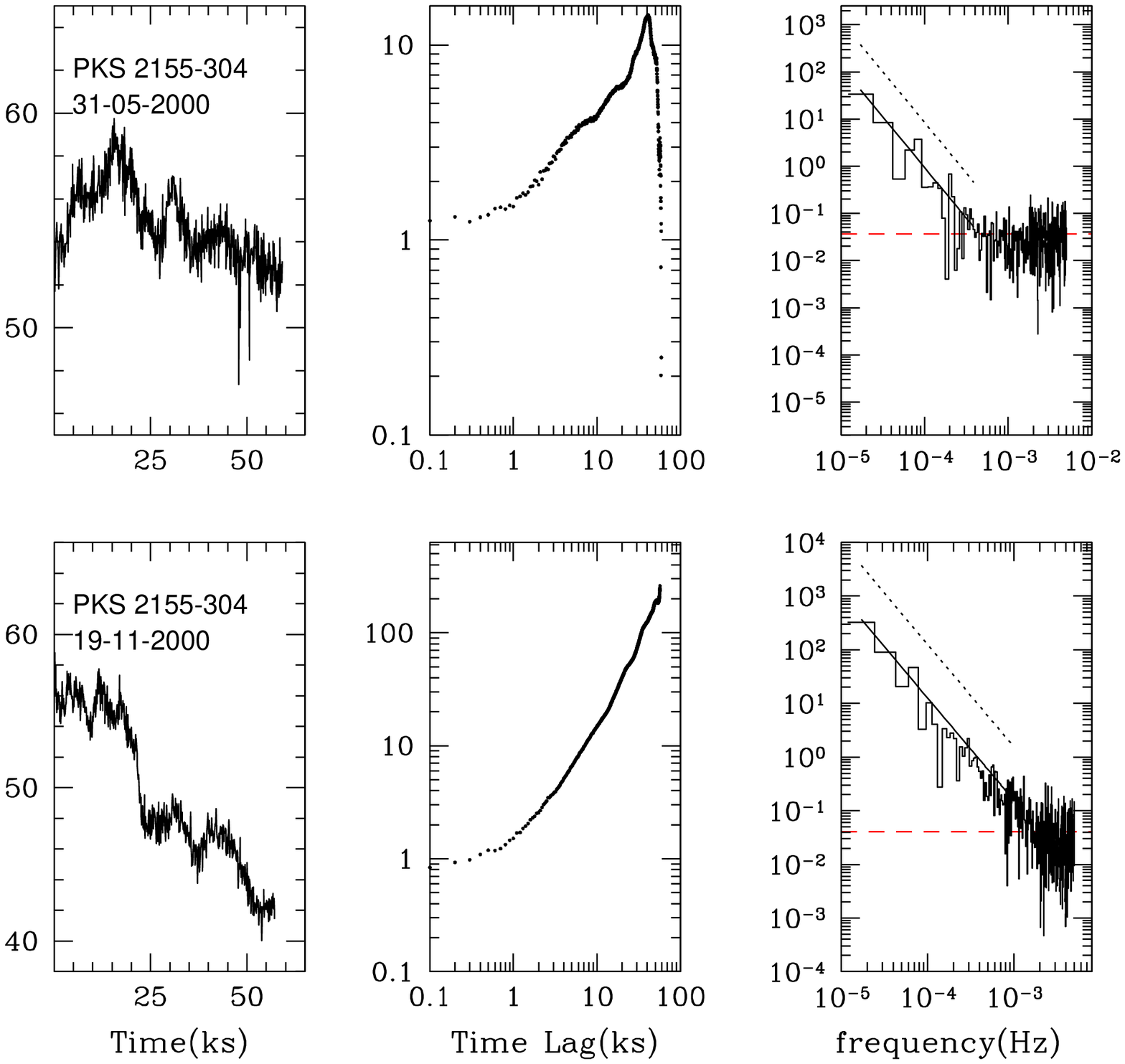}
\caption{As in Fig.\ 1 for 1ES 1426$+$428 and  PKS2155$-$304.}
\end{figure}

\clearpage

\begin{figure}
 \centering
  \includegraphics[width=6.0in,height=3.0in]{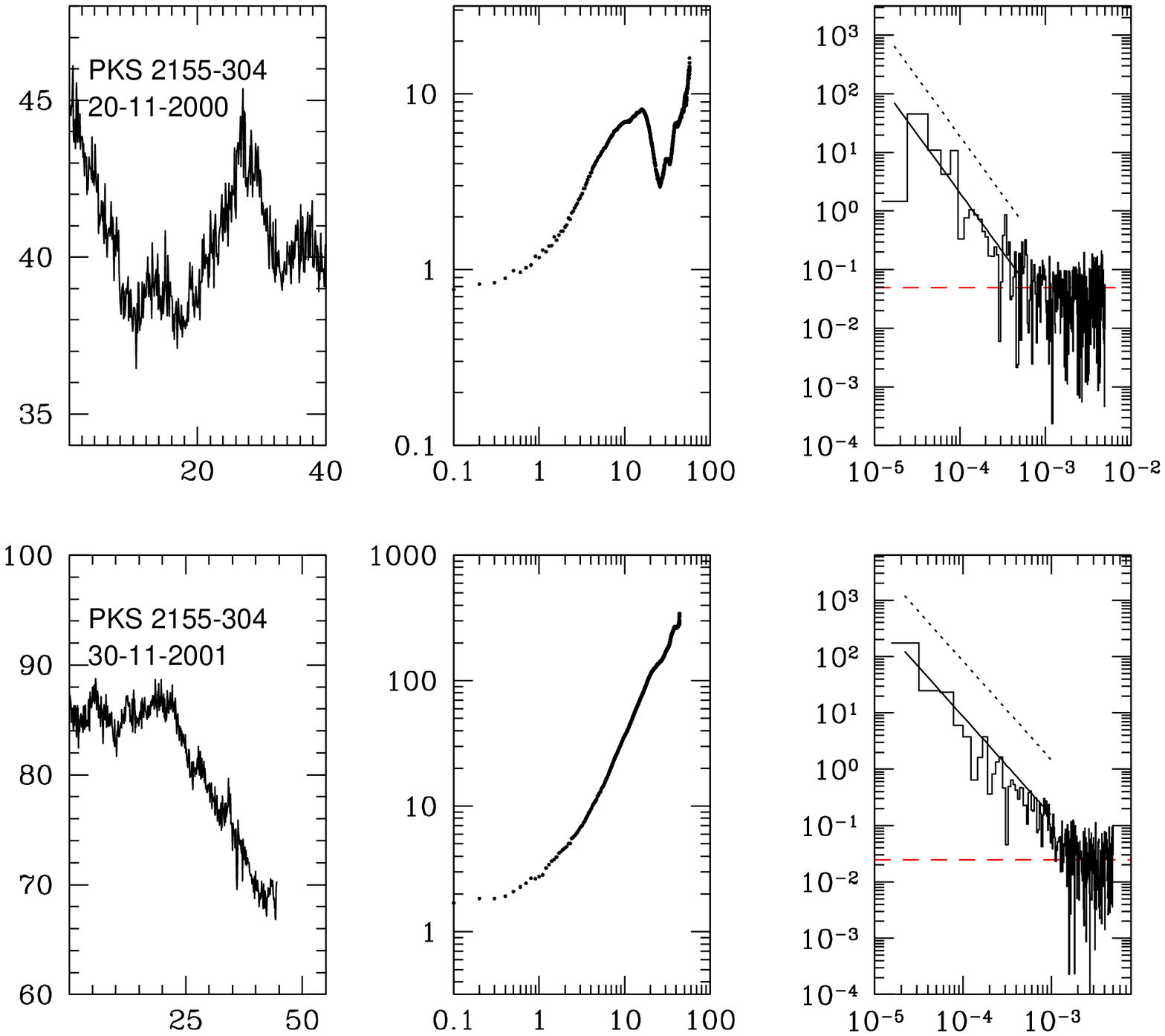}
  \includegraphics[width=6.0in,height=3.0in]{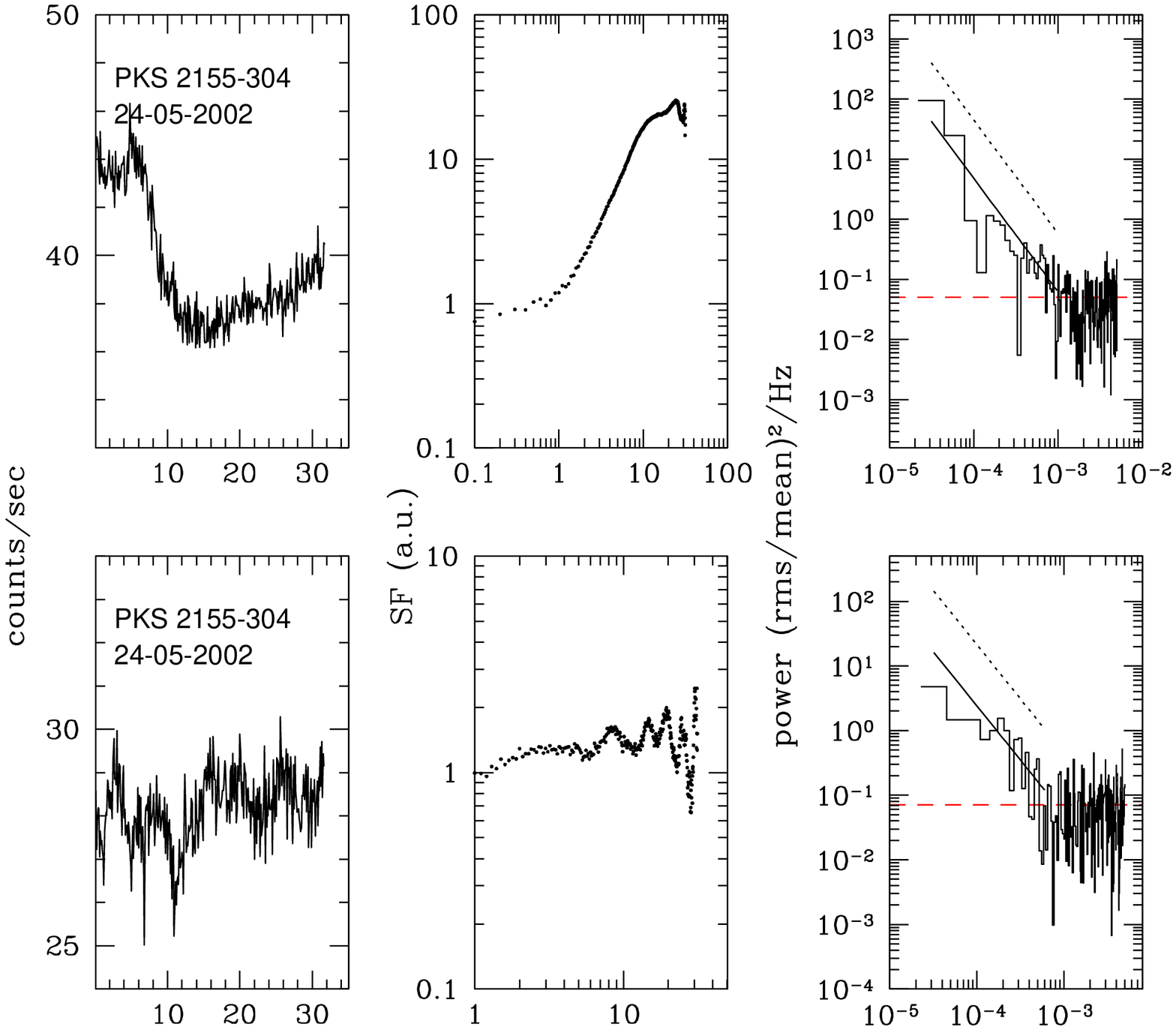}
  \includegraphics[width=6.0in,height=3.0in]{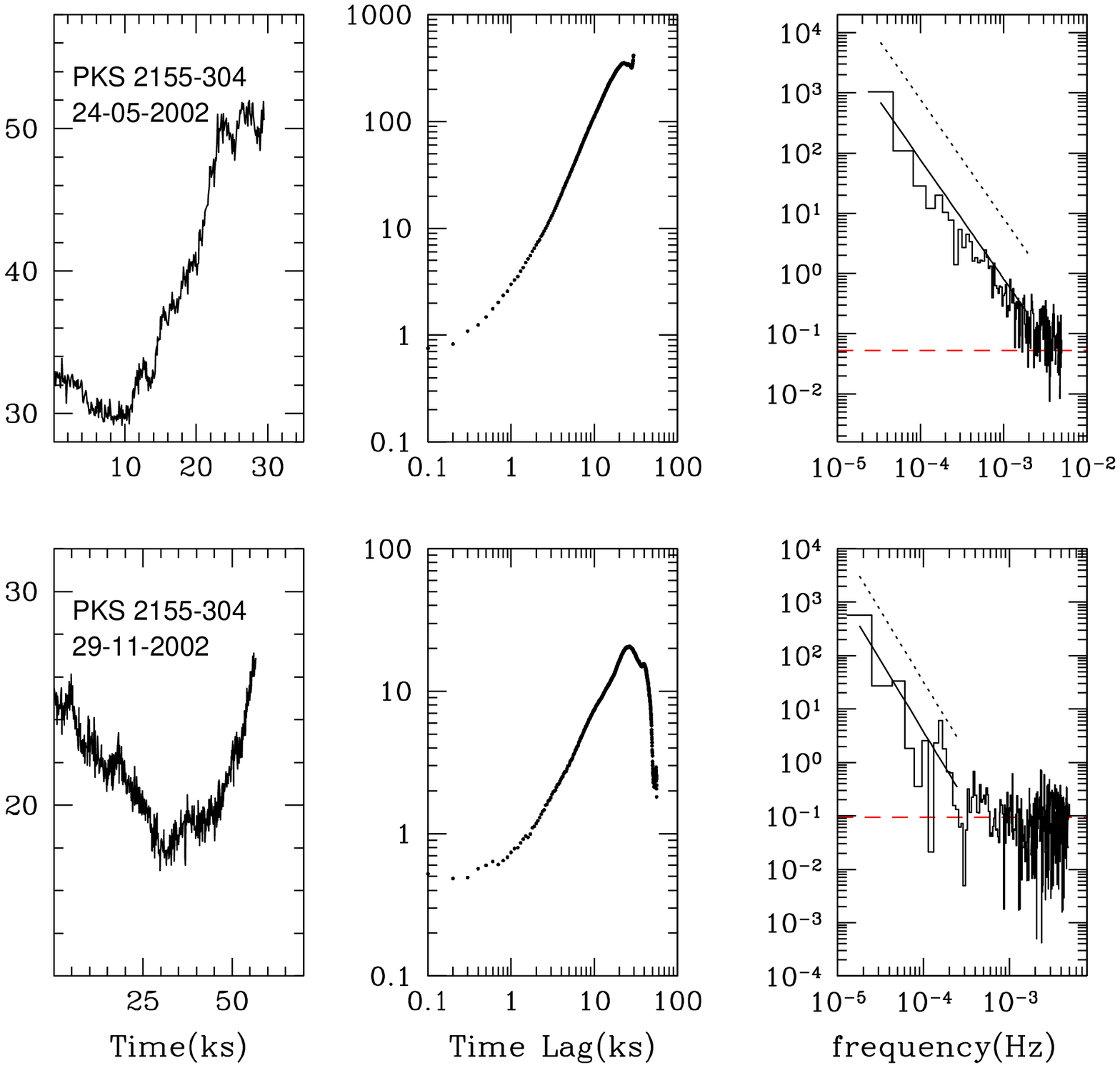}
\caption{As in Fig.\ 1 for  PKS2155$-$304.}
\end{figure}

\clearpage

\begin{figure}
 \centering
  \includegraphics[width=6.0in,height=3.0in]{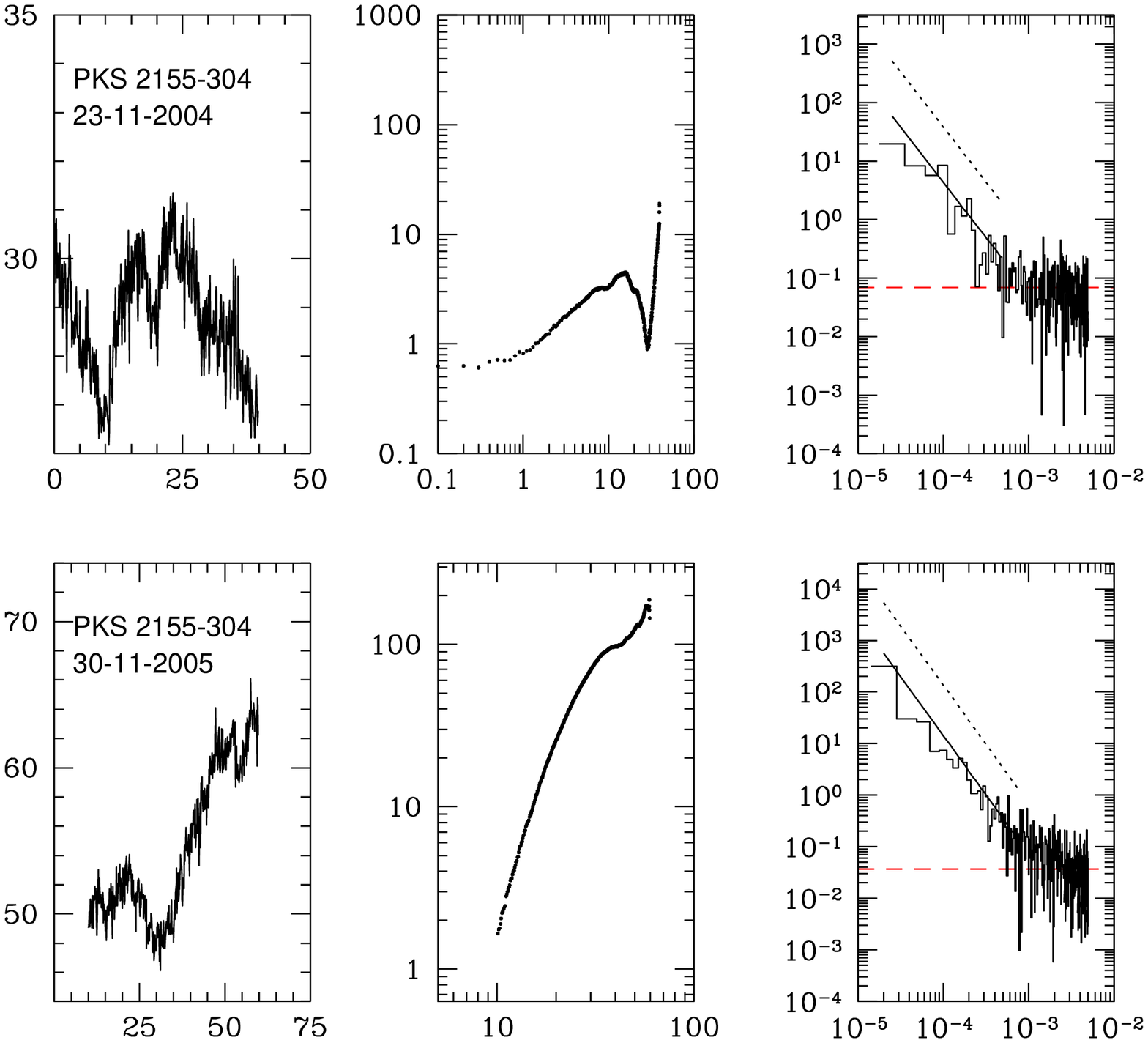}
  \includegraphics[width=6.0in,height=3.0in]{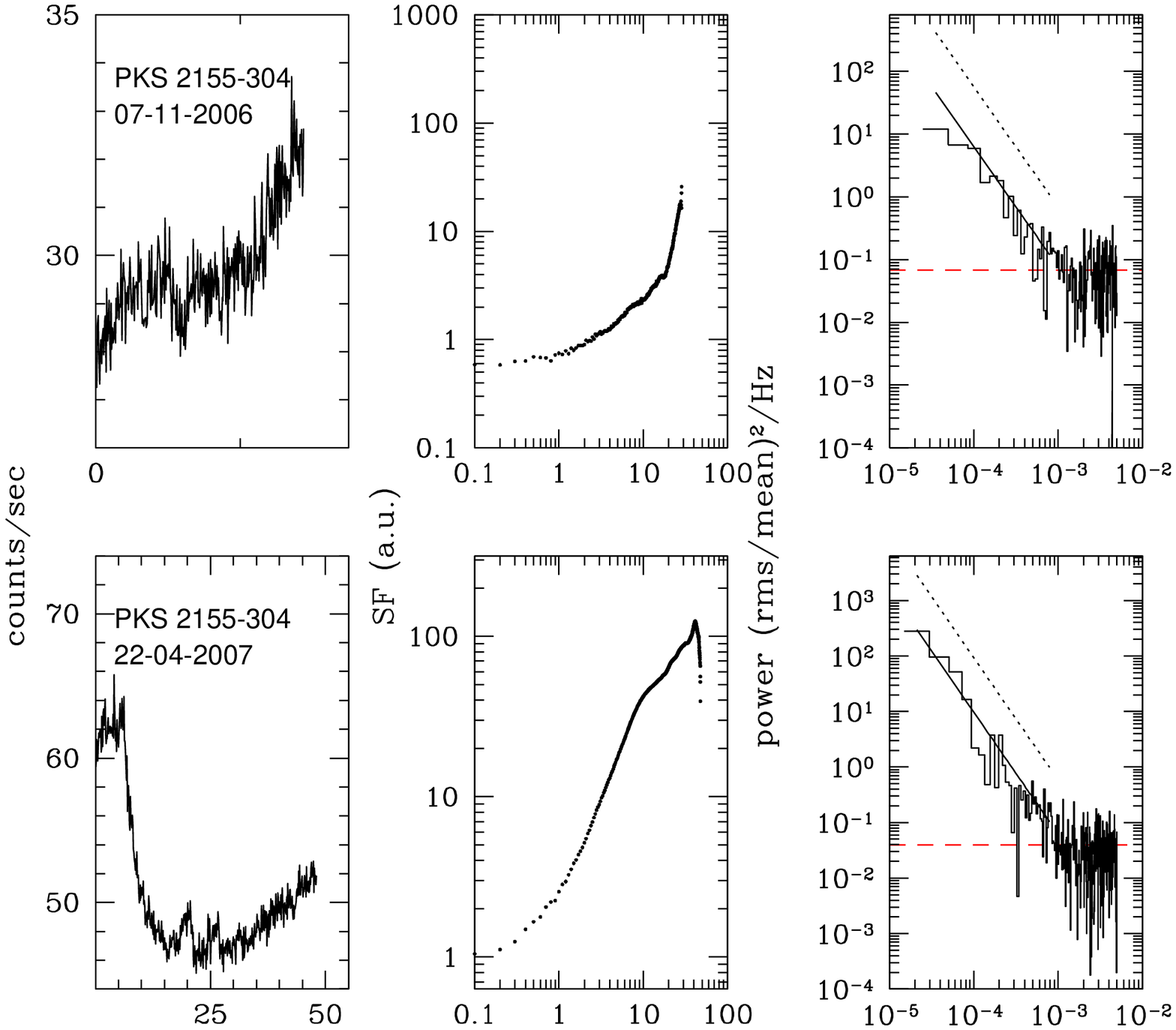}
  \includegraphics[width=6.0in,height=3.0in]{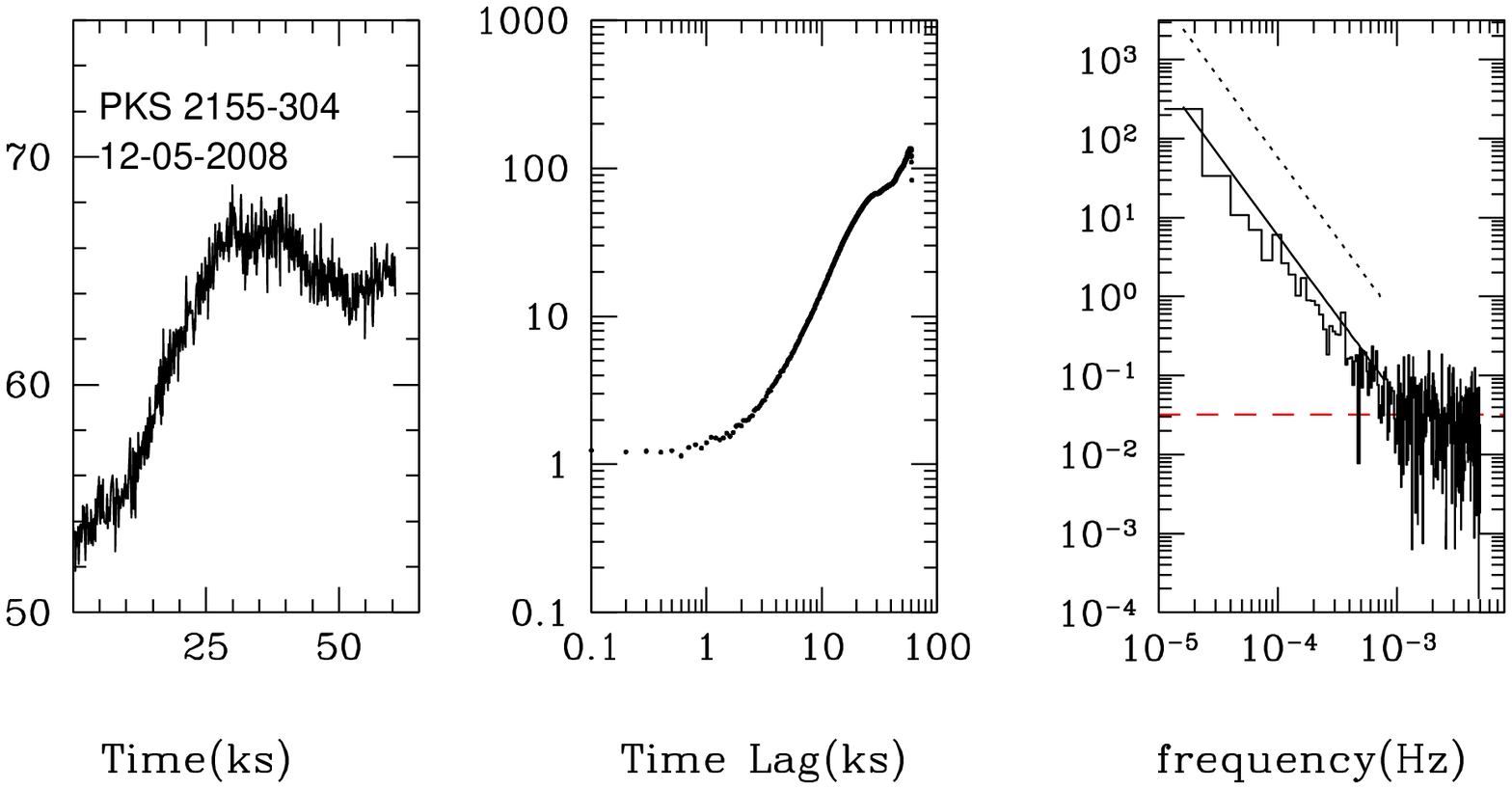}
\caption{As in Fig.\ 1 for  PKS2155$-$304.}
\end{figure}

\end{document}